\documentclass{elsart}
\usepackage{graphicx}

\begin{document}
\begin{frontmatter}
\title{An experimentally robust technique for halo measurement using the IPM at the Fermilab Booster}

\date{September 11, 2006}

\author[FNAL]{J. Amundson},
\ead{amundson@fnal.gov}
\author[FNAL]{W. Pellico},
\author[IIT]{L. Spentzouris},
\author[FNAL]{P. Spentzouris},
\ead{spentz@fnal.gov}
\author[FNAL]{T. Sullivan}
\address[FNAL]{Fermi National Accelerator Laboratory, Batavia, IL 60510, USA}
\address[IIT]{Department of Biological, Chemical, and Physical Sciences,
 Illinois Institute of Technology, Chicago, IL 60616, USA}

\begin{abstract}
We propose a model-independent quantity, $L/G$, to characterize
non-Gaussian tails in beam profiles observed with the Fermilab Booster
Ion Profile Monitor. This quantity can be considered a measure of beam
halo in the Booster. We use beam dynamics and detector simulations to
demonstrate that $L/G$ is superior to kurtosis as an experimental
measurement of beam halo when realistic beam shapes, detector effects
and uncertainties are taken into account. We include the rationale and
method of calculation for $L/G$ in addition to results of the
experimental studies in the Booster where we show that $L/G$ is a
useful halo discriminator.
\end{abstract}
\end{frontmatter}
\section{Introduction}

The generation and characterization of beam halo has been of increasing 
importance as beam intensities increase, particularly in a machine such as 
the Fermilab Booster.  The FNAL Booster provides a proton beam that 
ultimately is used for collider experiments, antiproton production, and 
neutrino experiments -- all high intensity applications.  The injection 
energy of the Booster is 400 MeV, low enough that space charge dynamics 
can contribute to beam halo production.  Beam loss, in large part due to 
halo particles, has resulted in high radiation loads in the Booster ring.  
The beam loss must be continuously monitored, with safeguards to insure 
that average loss rates do not exceed safe levels for ground water, the 
tunnel exterior, and residual radiation within the tunnel.  Collimation 
systems have been installed to localize beam loss within shielded 
regions~\cite{ref:shielding}.

Methods have been developed to
characterize beam halo that are based on analyzing the kurtosis of the beam
profile~\cite{ref:kurtosis1,ref:kurtosis2,ref:synergia}.  Kurtosis
is a measure of whether a data set is peaked or flat relative to a
normal (Gaussian) distribution.  Distributions with high kurtosis have
sharp peaks near the mean that come down rapidly to heavy tails.  The
kurtosis will decrease or go negative as a distribution becomes more
square, such as in cases where shoulders develop on the beam profile.
An important feature of quantifiers such as the {\it profile} and {\it
halo parameters} (introduced by Crandall, Allen, and Wangler in Refs.~\cite{ref:kurtosis1} and \cite{ref:kurtosis2}) is that
they are model independent.  Such beam shape quantities based on moments
of the beam particle distribution are discussed in Ref.~\cite{allwa},
which extends previous work based on 1D spatial
projections~\cite{ref:kurtosis1} to a 2D particle beam phase space
formalism.  These studies utilize numerical simulations of the physics
which cause the halo generation, but ignore the potential effects of
the detectors used in the experimental measurements. In realistic
accelerator operating conditions and, to some extent, also in
carefully prepared dedicated beam halo measuring experiments,
instrumental effects can reduce the effectiveness of the above
defined quantities.  Even when state-of-the-art 3D numerical
simulations are used to model realistic experiments, the
models fail to describe the measured beam distributions to the desired
detail~\cite{leda}.  In the following, we will investigate the use of shape
parameters relevant to the profiles observed with the Fermilab Booster
IPM, and will study how they perform using realistice 3D particle beam
simulations and a model of the IPM detector response.

Beam profile measurements in the Fermilab Booster are almost
exclusively done with an Ion Profile Monitor (IPM)~\cite{ref:ipm}.
The Booster IPM is able to extract horizontal and vertical beam profiles on a
turn-by-turn basis for an entire Booster cycle.  The IPM relies on the
ionization of background gas by the particle beam in the vicinity of
the detector.  A high voltage field is applied locally, causing the
ionized particles to migrate to micro-strip counters on a
multi-channel plate.  The applied high voltage is perpendicular to the
measurement axis of the multi-channel plate, so that it does not alter
the relative transverse positions of the ions.  However, the space
charge fields of the particle beam affect the transverse
trajectories of the ions, requiring a sophisticated calibration of the
IPM to relate the measured width of the distribution to the true width
of the particle beam~\cite{ref:ipmcalibration}.

This paper describes a new quantity for characterizing beam profiles,
$L/G$. We start by describing the observed beam profiles in the
Booster and their parametrization. We then define and calculate
kurtosis and $L/G$ for the resulting functional form. Next, we
consider the effect of the systematic and statistical uncertainties in
the IPM detector on both kurtosis and $L/G$. We then combine a
simulation of beam dynamics with a simulation of the IPM detector
response in order to demonstrate that $L/G$ can be used to
characterize the non-Gaussian beam tails in the Booster. We compare
the sensitivity of $L/G$ to that of kurtosis, and we find that $L/G$
is a superior discriminator of beam halo when realistic beam shapes,
detector effects and uncertainties are taken into account.  Finally,
we present the results of two Booster beam studies performed with and
without beam collimators, which demonstrate the sensitivity of $L/G$
to non-Gaussian beam tails.

\section{Booster beam profiles}
\begin{figure}[t]
\centering
\includegraphics*[width=0.7\columnwidth]{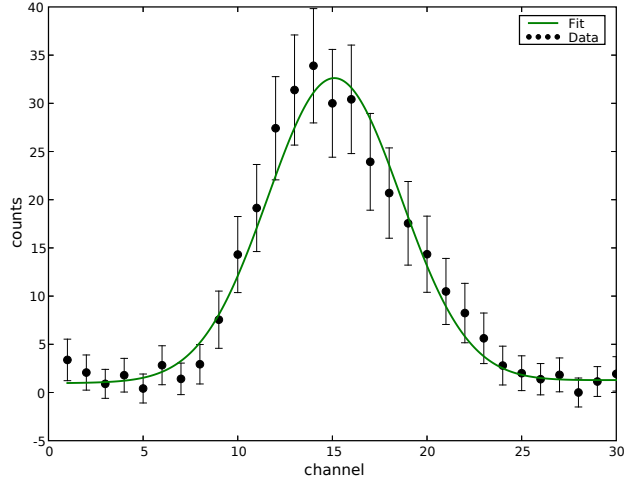}
\caption{Typical beam profile as observed using the Booster IPM. The fit is to the function $f(x)$ (Eq.~\ref{gauss_lin}).} 
\label{fig:fit}
\end{figure}

The Booster IPM is used to measure beam profiles under a wide range of operating conditions, ranging from normal operating conditions to machine tuning and beam studies under potentially extreme conditions. The beams under these conditions vary considerably. Following the standard experimental procedure for characterizing a peak signal combined with a potentially large background by fitting to a Gaussian plus polynomial background, the IPM data are characterized by fitting the profiles to a sum of Gaussian and linear functions~\cite{ref:ipm},
\begin{equation}
	f(x)=Ng(x)+Ml(x), \label{gauss_lin}
\end{equation}
where
\begin{equation}
	g(x)=\frac{1}{\sqrt{2\pi}\sigma}\exp\left(-\frac{(x-x_{0})^{2}}{2\sigma^{2}}\right),
\end{equation}
\begin{equation}
	l(x)=1+c_{1}(x-x_{0}),
\end{equation}
and $N$, $M$ ,$\sigma$, $x_{0}$, and $c_{1}$ are the fitting parameters.
This parametrization does a reasonable job of characterizing the
observed IPM profiles. Fig.~\ref{fig:fit} shows a typical beam profile observed during normal Booster operations, along with the fit to Eq.~\ref{gauss_lin}.

\section{Characterizing observed profiles}
Our goal is to describe the shape of the observed beam profiles by some parameter derived from the observed beam shape. For simplicity, we start by ignoring all detector effects and assume that the measured beam shape is well described by the function $f(x)$ defined above.

A standard technique for characterizing the shape of a distribution is to calculate the kurtosis of the distribution, $k$, defined by
\begin{equation}
k\equiv\frac{\langle(x-x_{0})^{4}\rangle}{\langle(x-x_{0})^{2}\rangle^{2}}-3.
\end{equation}

We now calculate the kurtosis of $f(x)$ (Eq.~\ref{gauss_lin}). Since the function $l(x)$ is not compact, we have to restrict ourselves
to a fixed range in $x$. We take
the region $x_{0}\pm n\sigma$ and use $n=5$ wherever numerical results
are needed. Ignoring the irrelevant overall normalization, we set
\begin{equation}
M=\frac{\mu}{2n\sigma}
\end{equation}
and
\begin{equation}
N=\frac{1-\mu}{{\rm erf} \left(n/\sqrt{2}\right)},
\end{equation}
where we have introduced the parameter $\mu$ to characterize the relative fractions of Gaussian and linear components. With this definition
\begin{equation}
\int_{-n\sigma}^{+n\sigma}Ml(x) \d x=\mu
\end{equation}
and
\begin{equation}
	\int_{-n\sigma}^{+n\sigma}Ng(x)\d x=1-\mu.
\end{equation}

\begin{figure}[t]
 \centering
 \includegraphics*[height=0.7\columnwidth,angle=270]{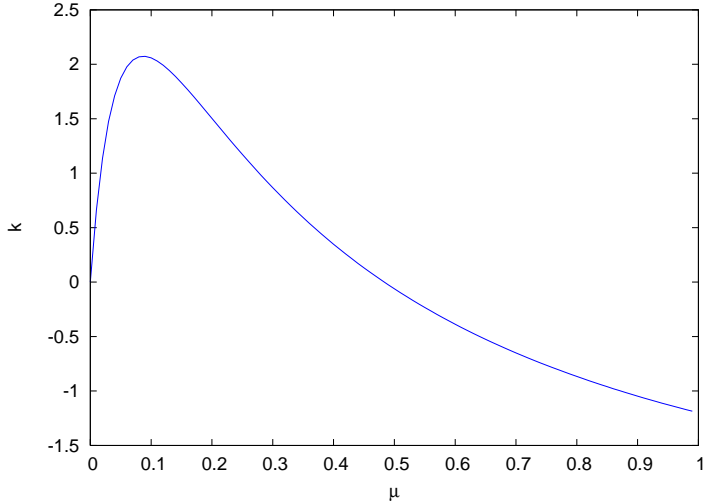}
 \caption{Kurtosis as a function of the parameter $\mu$ for $n=5$.}
 \label{fig:kurtosistheory}
\end{figure}

We now evaluate
\begin{equation}
 k(\mu) = \frac{\int_{-n\sigma}^{+n\sigma}(x-x_0)^4f(x) \d x}%
{\left(\int_{-n\sigma}^{+n\sigma}(x-x_0)^2f(x) \d x\right)^2}.
\end{equation}
Neglecting the small difference
$1-{\rm erf}\left(n/\sqrt{2}\right)\approx6\cdot10^{-7}$ for $n=5$, we obtain
\begin{equation}
k(\mu)=\frac{9\mu n^4-135 \mu+135}{5 \mu^2 n^4+\left(30 \mu
-30 \mu^2\right) n^2+45 \mu^2-90 \mu+45}-3,%
\label{eq:kmu}
\end{equation}
which is shown in Fig.~\ref{fig:kurtosistheory}.

Our goal is to use an experimentally accessible observable to quantify the non-Gaussian portion of the observed beam, which amounts to using the inverse of Eq.~\ref{eq:kmu}. From Fig.~\ref{fig:kurtosistheory} it is clear that kurtosis is not a good observable for this problem. In order to extract beam shape information $\mu$ from an observable $k$, we need the inverse function $\mu(k)$. For this beam shape, $\mu(k)$ is multi-valued for a significant portion of the possible range of $\mu$. A beam that is roughly half non-Gaussian ($\mu\sim0.5$) has the same kurtosis as a beam with no non-Gaussian component whatsoever ($\mu=0$). At best, kurtosis gives a qualitative measure of the degree to which the distribution is non-Gaussian.

\begin{figure}[t]
\centering
\includegraphics*[width=0.7\columnwidth]{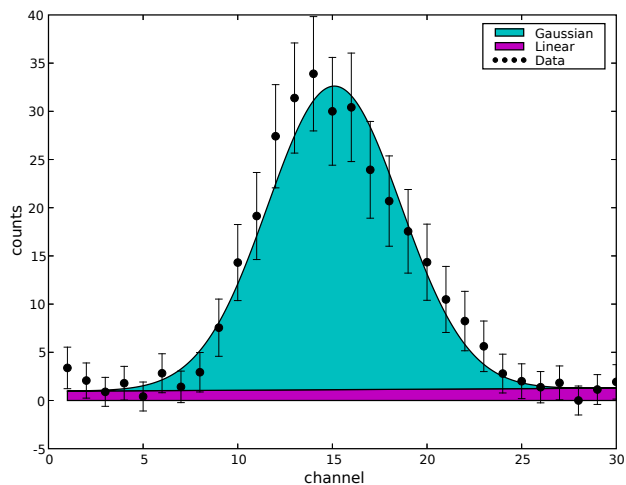}
\caption{Fitted IPM profile from Figure~\ref{fig:fit} showing Gaussian and linear (non-Gaussian) contributions in cyan and magenta, respectively.} 
\label{fig:portions}
\end{figure}

In order to find a more quantitative shape measure that is compatible with the experimentally observed Gaussian plus linear shape, we start by defining the integral quantities $L$ and $G$ by
\begin{equation}
	L \equiv \int_{-n\sigma}^{+n\sigma}Ml(x) \d x
\end{equation}
and
\begin{equation}
	G \equiv \int_{-n\sigma}^{+n\sigma}Ng(x) \d x
\end{equation}
and use the ratio $L/G$ as our new shape parameter. The ratio has an
obvious geometrical interpretation: it is the ratio of areas of the
non-Gaussian and Gaussian portions of the beam profile over the range
$\pm n\sigma$. (See Fig.~\ref{fig:portions}.) As stated above, we use $n=5$ throughout this paper wherever numerical results are required. However, we always fit the beam profile using the entire range of the detector, regardless of $n$. The extracted value of $L/G$
includes all of the information in the fit -- a given value of $n$
is just a convention for interpreting the results\footnote{See the further discussion in Sec.\ref{sec:simulation}.}.

\begin{figure}[t]
 \centering
 \includegraphics*[height=0.7\columnwidth,,angle=270]{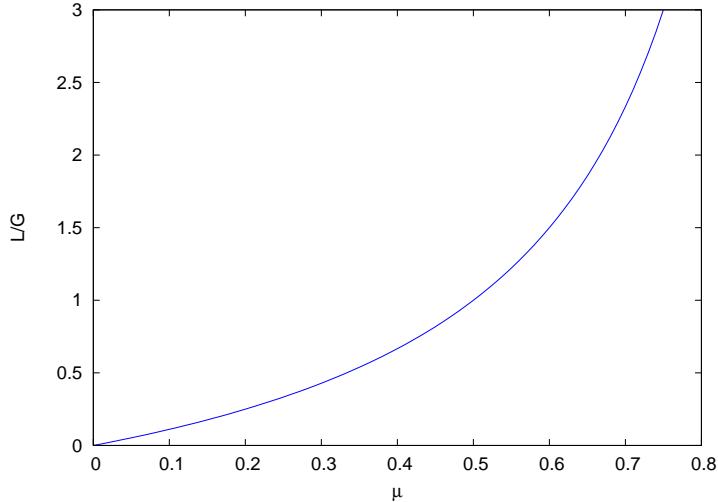}
 \caption{$L/G$ as a function of the parameter $\mu$.}
 \label{fig:lgtheory}
\end{figure}

Using the definitions of $N$, $M$, and $\mu$ above, we obtain
\begin{equation}
 	L/G~(\mu) = \frac{\mu}{(1-\mu)}
\end{equation}
which is shown in Fig.~\ref{fig:lgtheory}. 
$L/G$ has the advantage of being a monotonically increasing function of $\mu$, allowing for the unambiguous extraction of the inverse function $\mu(L/G)$, which has the simple form
\begin{equation}
	\mu(L/G) = \frac{L/G}{1+L/G} .
\end{equation}

\section{Detector effects}
Extracting beam shape information from the Booster IPM is complicated by the beam intensity-dependent smearing of the signal and statistical noise. The smearing effect comes from the electromagnetic field generated by the beam itself. The observed beam shape is a convolution of the true shape with the response function of the detector, the {\em
smearing function}. The authors of Ref.~\cite{ref:ipmcalibration}
studied the IPM smearing function and developed a system for
extracting the true beam width from the observed IPM data. In the case
discussed in the reference, the objective was to study the evolution
of the statistical emittance of the beam, which only depends on the
second moment of the beam profile. By assuming a Gaussian shape for
the true beam distribution, it was possible to invert the smearing
function as it applies to the beam size. Experimental measurements utilizing independent measurements of the beam size verified that the inversion of the smearing function is accurate when applied to beam size. 

Since halo studies require quantitative measurement of the
{\em shape} of the beam in addition to the size, any inversion of the smearing function would be strongly dependent on the assumed shape. Furthermore, independent measurements of the true beam shape are not readily available, so experimental verification is excluded.
The method we present in these paper avoids the difficulties of
inverting the IPM smearing function by operating directly on the raw
IPM data. Where we compare the experimental data with predictions from
beam dynamics simulations, we smear the simulation results in order to
compare them with data.

The smearing function is not the only experimental complication. Each channel in the IPM includes a constant pedestal that must be subtracted from the data.
Before each IPM run, 30 pedestal
triggers are collected; these triggers are used to find the mean and
standard deviation of the pedestal.  
\begin{figure}[t]
\centering
\includegraphics*[width=0.7\columnwidth]{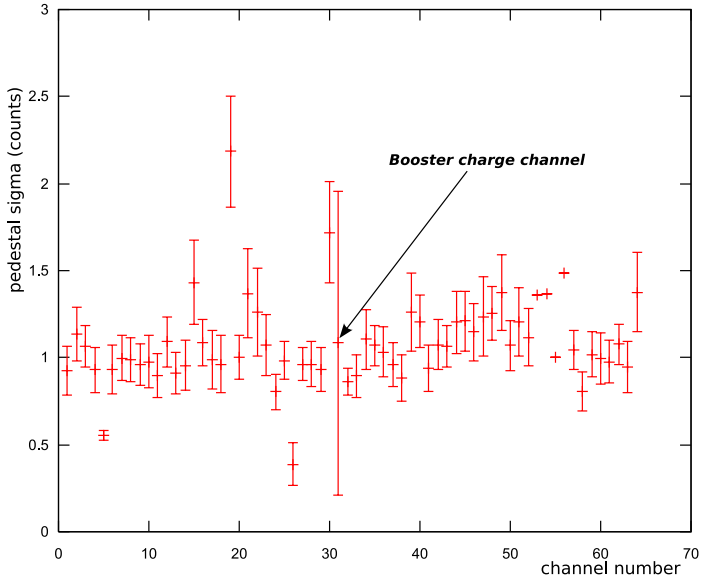}
\caption{Average standard deviation (sigma) of the pedestal of each IPM 
channel versus channel number.  The standard deviation for each pedestal 
was averaged for 35 consecutive runs (Booster cycles); the error bars shown
represent the RMS of these 35 runs.  Note that channel 31 is used to
measure the Booster charge as a function of turn, thus it is natural
that the characteristic spread of each pedestal is different than the
rest of the channels.} 
\label{fig:peds}
\end{figure}
Fig.~\ref{fig:peds} shows the typical pedestal variation present in the IPM. The pedestal introduces a fluctuation in the data that is constant across the entire detector. We take this source of error into account when calculating the fits to Eq.~\ref{gauss_lin}, so the expected effect on $L/G$ is only a small increase in the overall error. The kurtosis calculation suffers a much larger error because of the effect of the $(x-x_0)^4$ term; the error in $x$ far from $x_0$ is magnified fourfold. We will see below that the end result is a smaller overall fractional error in $L/G$ than in kurtosis when extracting from data.

\section{Results using simulated data}\label{sec:simulation}

In order to test our $L/G$ technique and compare it with kurtosis, we employ an
integrated simulation of both the Booster accelerator and the IPM
detector. Our accelerator model is based on the code
Synergia~\cite{ref:synergia}, which includes halo generating effects such as
nonlinearities in the accelerator lattice and 3D space-charge effects.
Our IPM detector model includes both the space-charge effects on the
IPM response~\cite{ref:ipmcalibration} and pedestal variations.  We
simulate a variety of initial beam shapes, ranging from a Gaussian
with no space-charge effects, to beam with severe halo component, much
larger than what is usually observed experimentally at the Booster.  In order to generate beams with large halo that are realistic, we construct an initial beam containing a matched Gaussian component combined with a completely flat (and mismatched) halo component, then use Synergia to simulate its evolution through two revolutions through the Booster. The resulting beam has a profile that is qualitatively consistent with observed beam profiles.

\begin{figure}[t]
\centering
\includegraphics*[height=0.5\columnwidth,,angle=270]{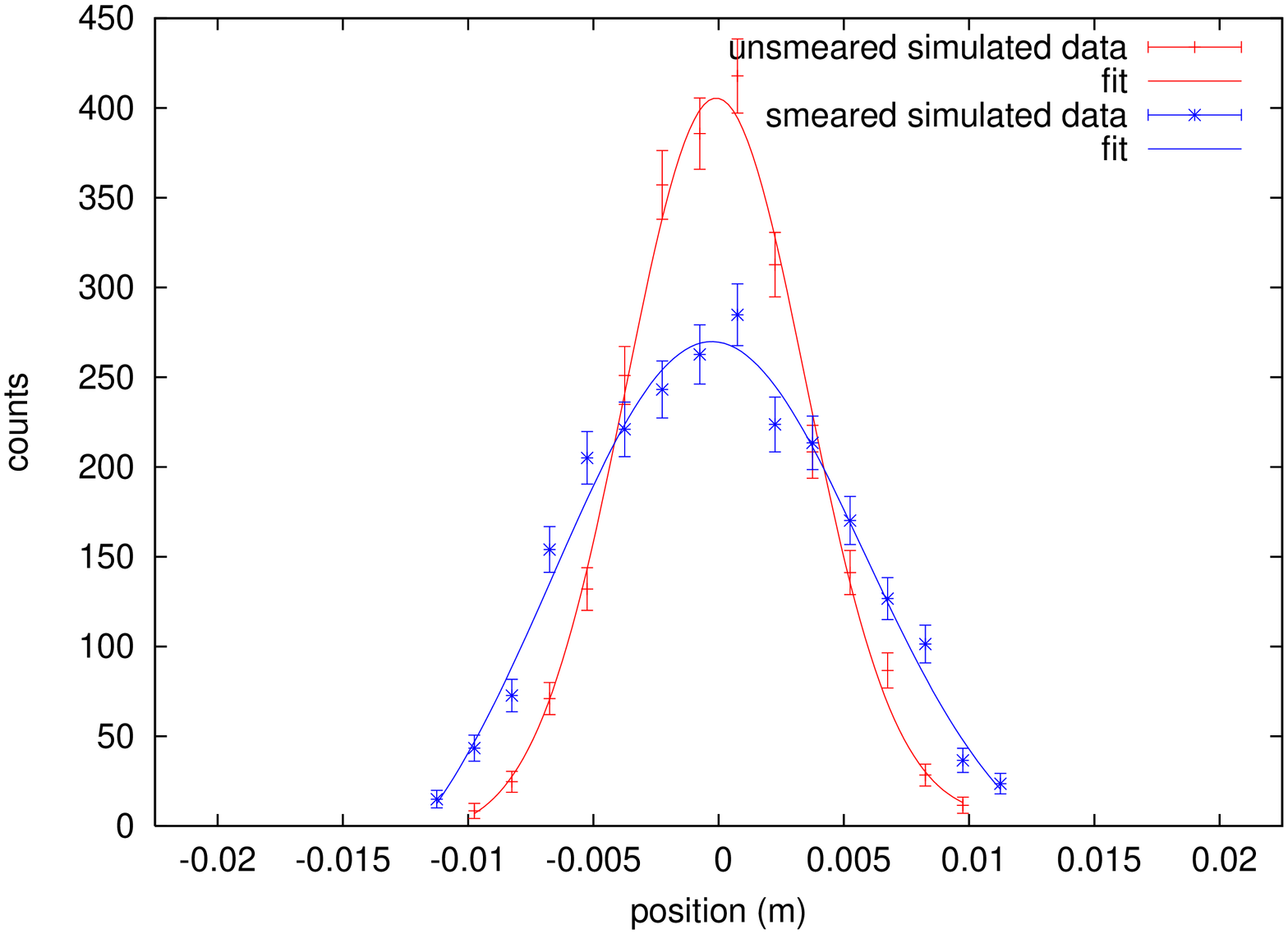}\includegraphics*[height=0.5\columnwidth,,angle=270]{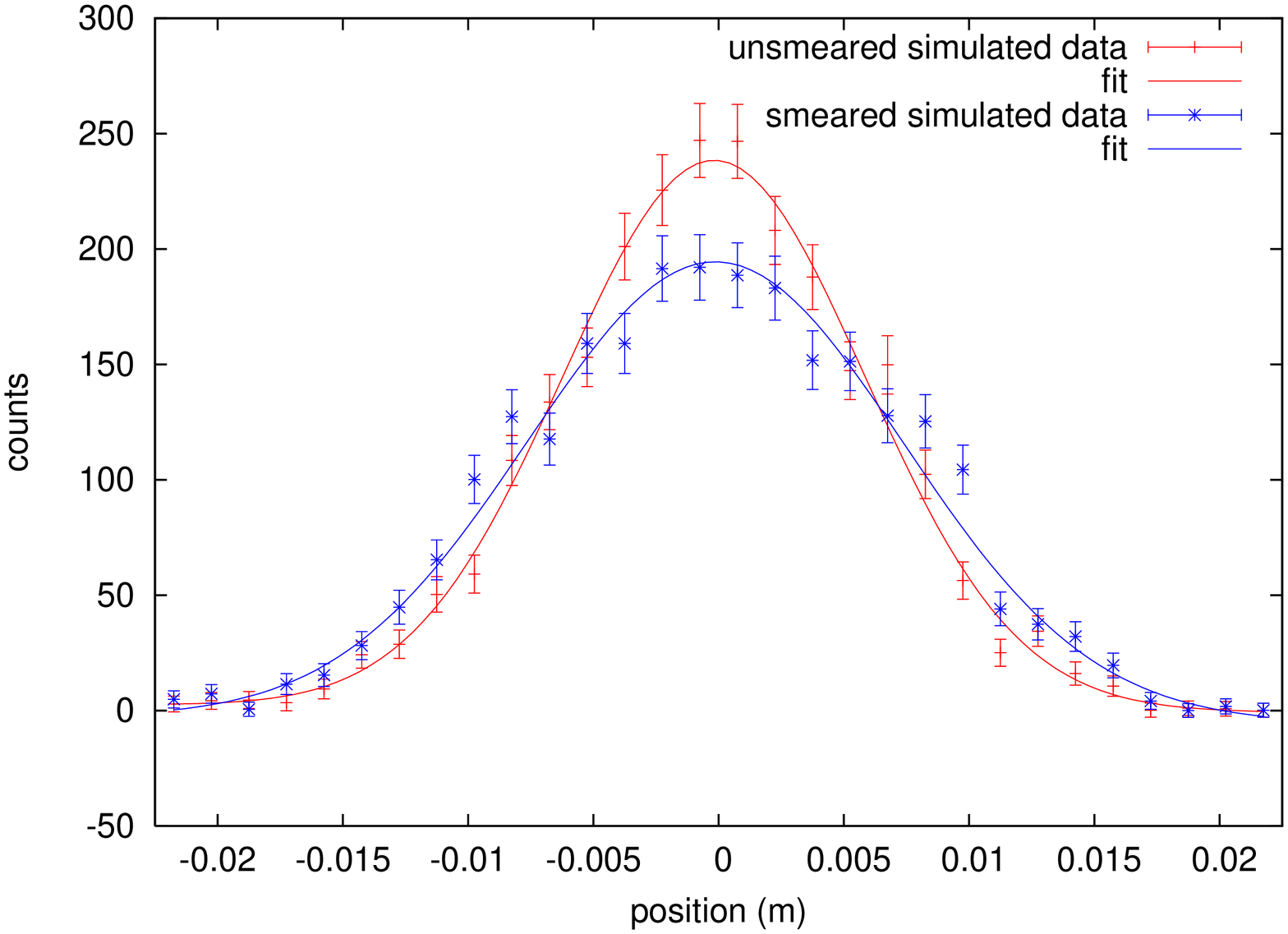}
\caption{Smeared and unsmeared simulated IPM beam profiles for a Gaussian beam.  The plot on the left shows a simulated profile with width similar to a typical measured horizontal width, the plot on the right has a width similar to a typical measured vertical width. } 
\label{fig:simu1}
\end{figure}

\begin{figure}[t]
\centering
\includegraphics*[height=0.5\columnwidth,,angle=270]{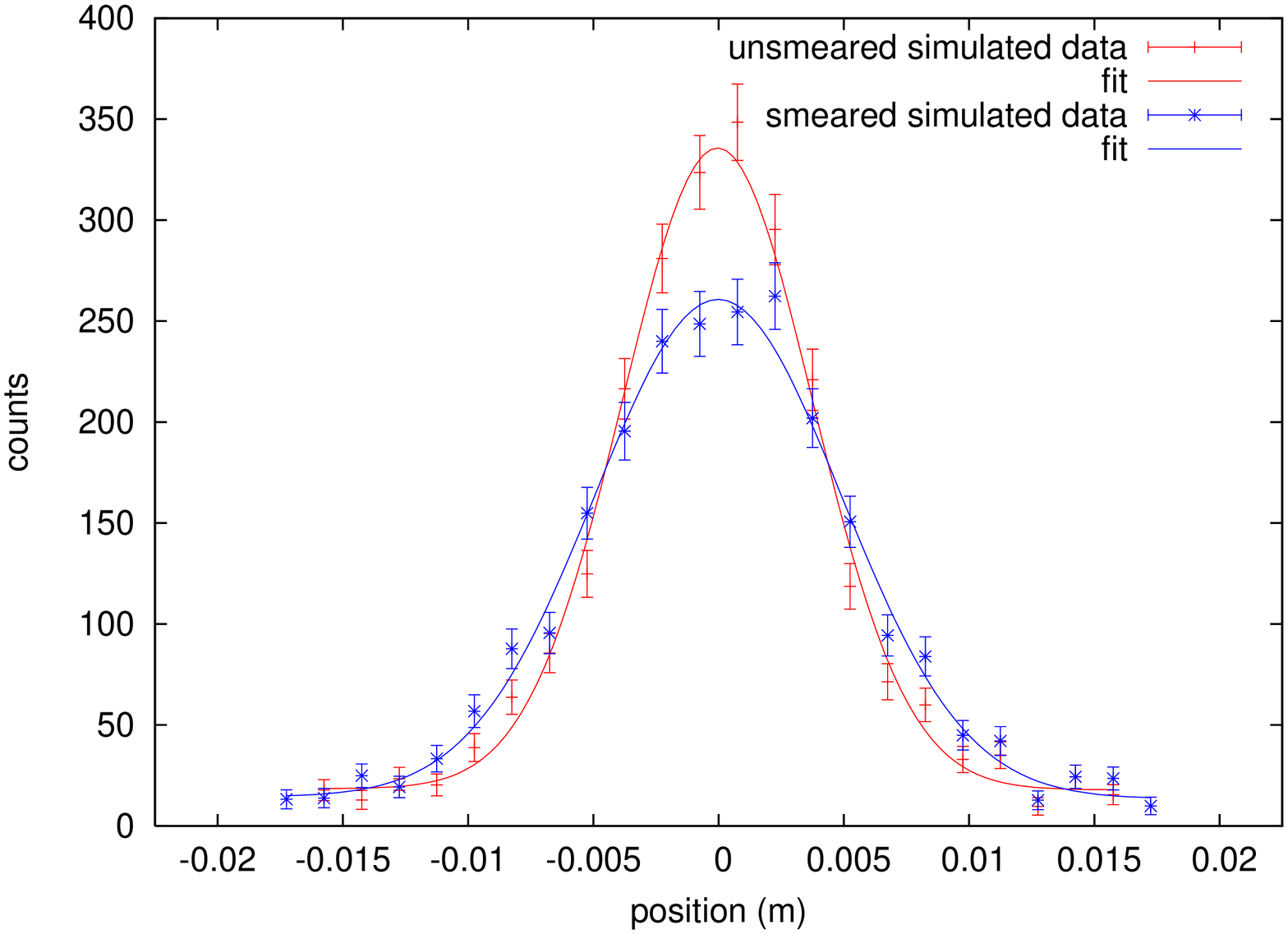}\includegraphics*[height=0.5\columnwidth,,angle=270]{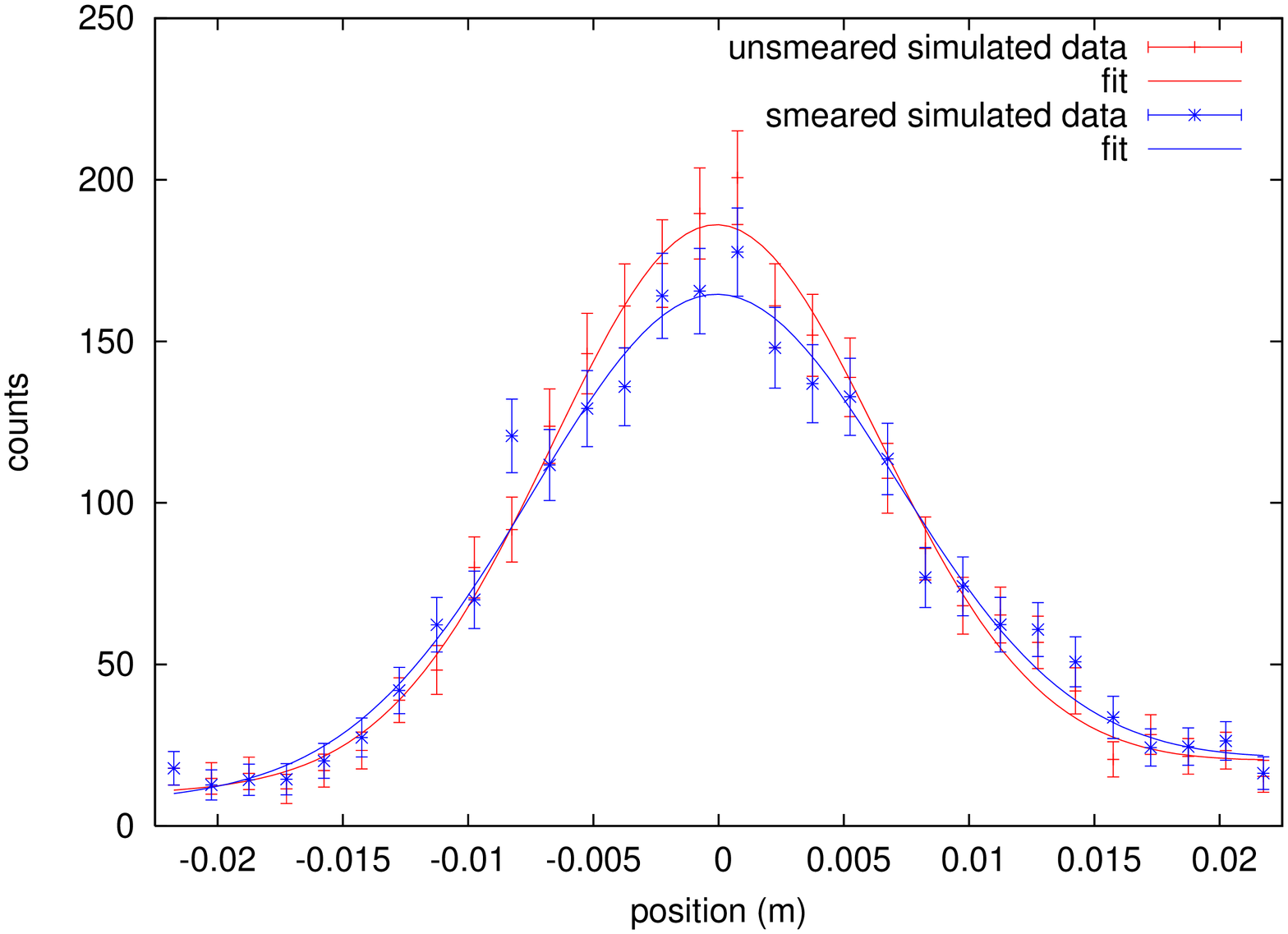}
\caption{Smeared and unsmeared simulated IPM beam profiles for a beam with a profile distribution that has tails similar to a typical Booster IPM profile. The plot on the left shows a simulated profile with width similar to a typical measured Booster IPM horizontal width, the plot on the right has a width similar to a typical measured Booster IPM vertical width.} 
\label{fig:simu2}
\end{figure}

\begin{figure}[t]
\centering
\includegraphics*[height=0.5\columnwidth,,angle=270]{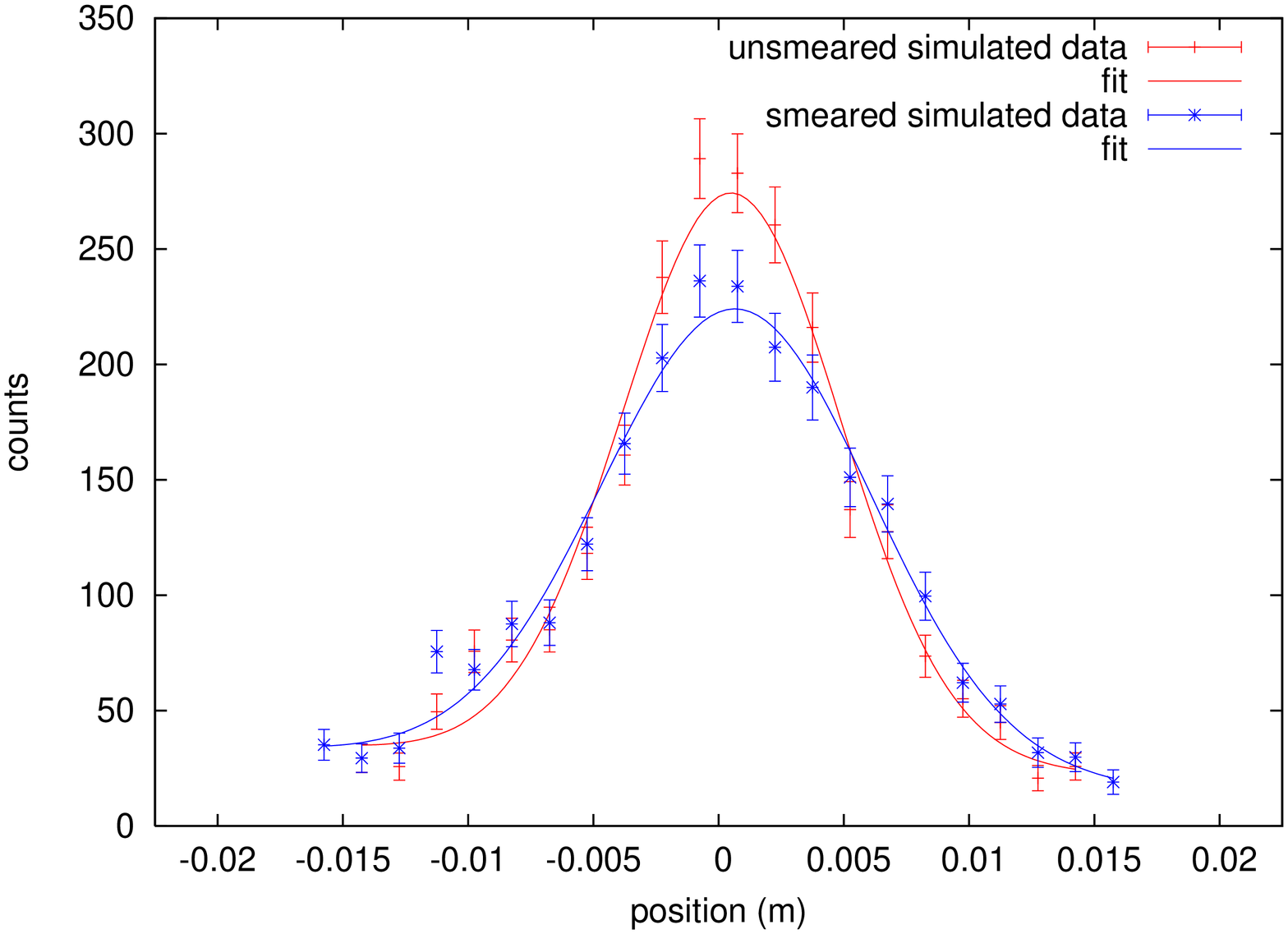}\includegraphics*[height=0.5\columnwidth,,angle=270]{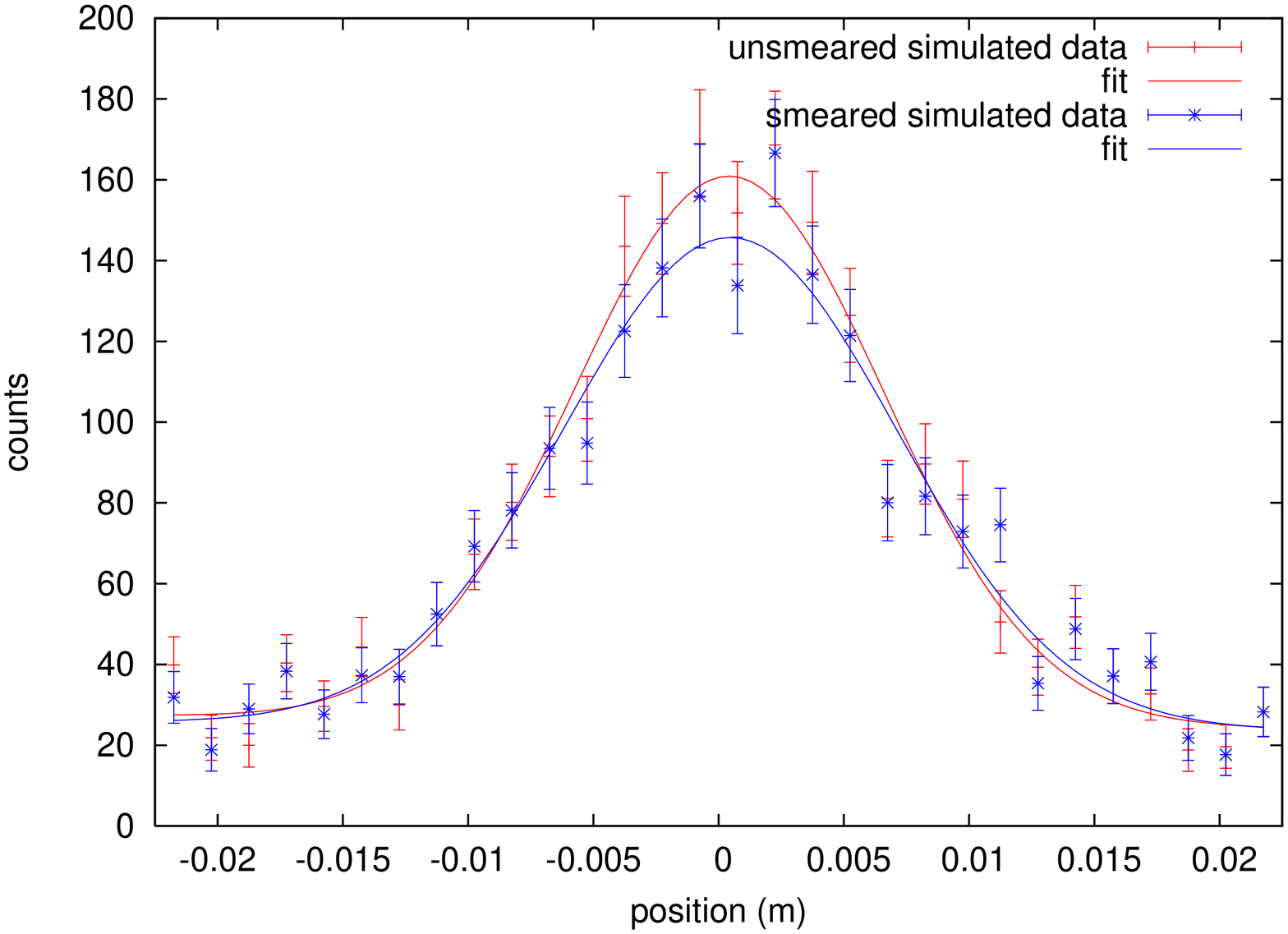}
\caption{Smeared and unsmeared simulated IPM beam profiles for a beam with large simulated halo component.  The plot on the left shows a simulated profile with width similar to a typical measured Booster IPM horizontal width, the plot on the right has a width similar to a typical measured Booster IPM vertical width.} 
\label{fig:simu3}
\end{figure}

\begin{figure}[t]
\centering
\includegraphics*[height=0.5\columnwidth,,angle=270]{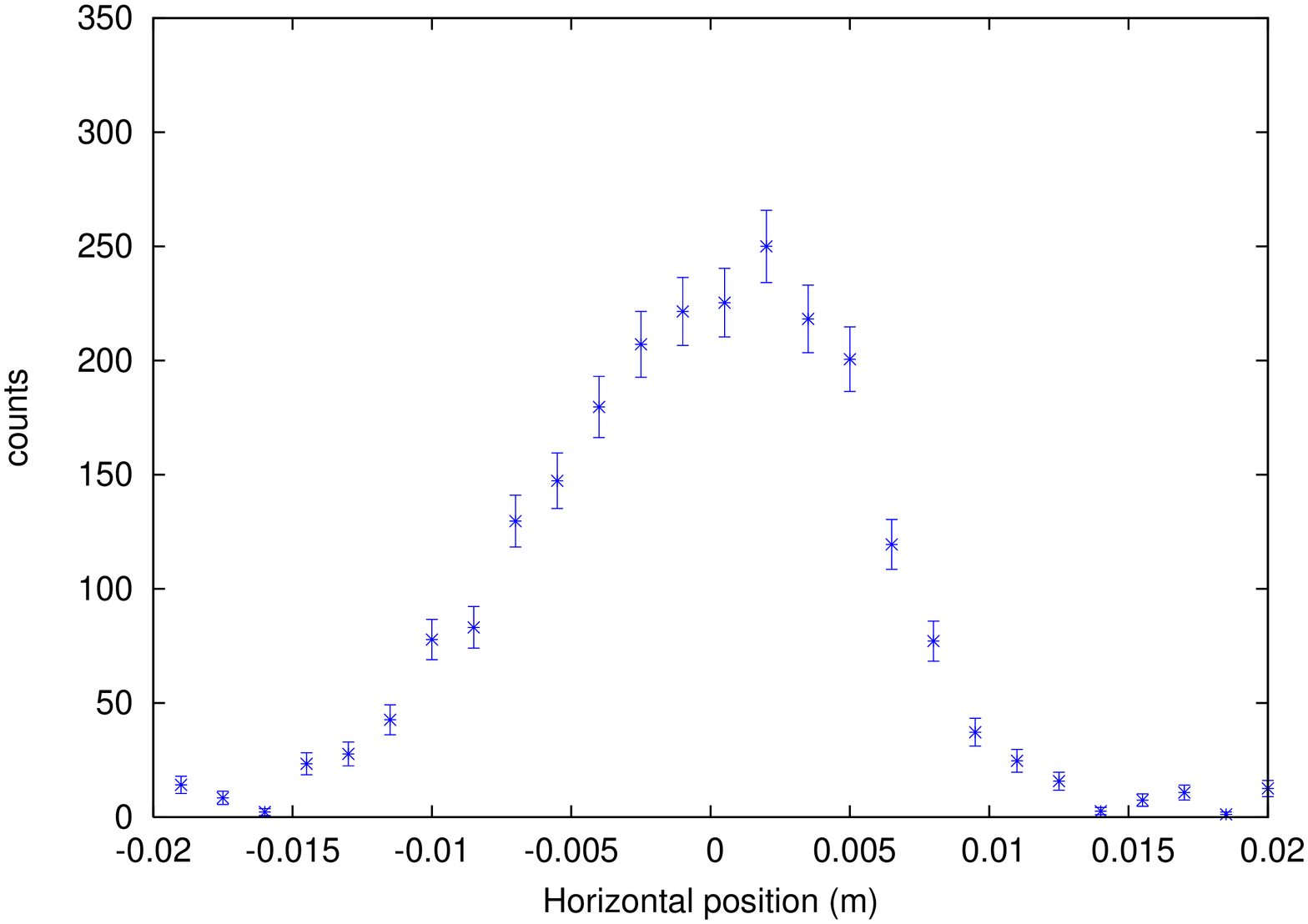}\includegraphics*[height=0.5\columnwidth,,angle=270]{smeared_x0.600000.eps}
\caption{Measured (left) and simulated (right), smeared and unsmeared, IPM beam profiles for a beam with large simulated halo component.} 
\label{fig:comp1}
\end{figure}

In Fig.~\ref{fig:simu1} we show simulated beam profiles using a purely Gaussian beam, with and without detector smearing effects. Figs.~\ref{fig:simu2} and \ref{fig:simu3} show profiles for a beam that has been adjusted to have a non-Gaussian component consistent with typical profiles observed in the Booster (Fig.~\ref{fig:simu2}) and the sort of beam observed under extremely mismatched conditions (Fig.~\ref{fig:simu3}). The fitted curves in these figures are fits to $f(x)$ (Eq.~\ref{gauss_lin}.) In Fig.~\ref{fig:comp1} we
show the measured profile for ten consecutive turns (added together to
increase statistics), obtained during beam study time with the Booster
operating with a severely mismatched beam, and the simulated (beam and
detector response) profiles for similar beam conditions: the agreement
is very good\footnote{The channels with zero counts in the data are
dead; this effect was not included in the simulation.}

Having established that we can simulate the beam and detector well enough to produce profiles qualitatively similar to the observed profiles, we are now ready to calculate kurtosis and $L/G$ for a controlled set of data. For input, we take seven different input beams with mixing parameters $\mu$ ranging from 0 to 0.7. We then pass them through two simulated revolutions of the Booster cycle and then the detector simulation. Although the input beams have an unrealistically large non-Gaussian component, by the time the beam enters the detector simulation, after a few turns through the accelerator simulation, the halo fraction has reached realistic levels.
 
\begin{figure}[t]
\centering
\includegraphics*[height=0.5\columnwidth,angle=270]{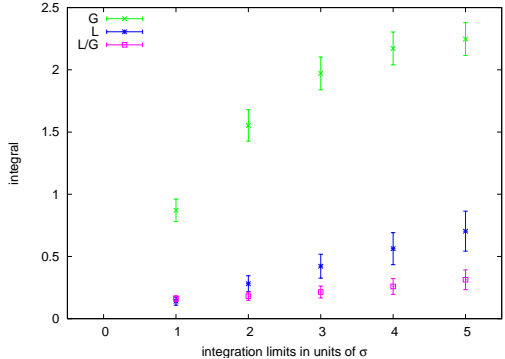}
\caption{Values of the $L/G$ components and $L/G$ as a function of the integration limits in units of the sigma of the Gaussian component.} 
\label{fig:components}
\end{figure}

As described above, $L/G$ depends on the parameter $n$, the number of units of $\sigma$ over which to perform our integrations. We have (somewhat arbitrarily) chosen $n=5$ for our results. In Fig.\ref{fig:components}, we show the variation of $L$, $G$ and the ratio $L/G$ as a function of $n$ for a simulated beam profile with a moderate non-Gaussian contribution.  The fits we always use all the information available to us, {\em i.e.}, the full range of the detector, so the choice of $n$ is an arbitrary convention.

\begin{figure}[t]
\centering
\includegraphics*[height=0.5\columnwidth,angle=270]{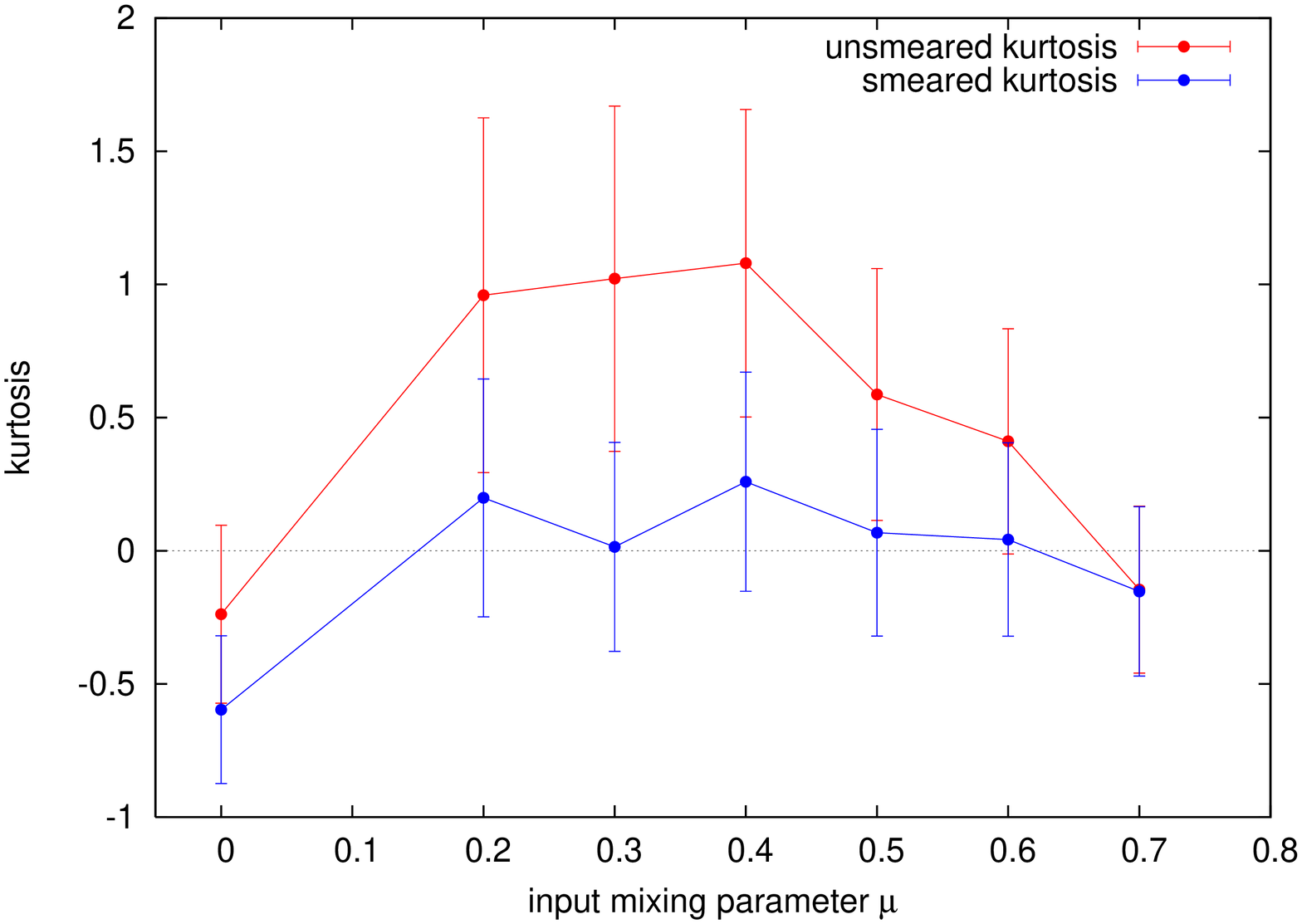}%
\includegraphics*[height=0.5\columnwidth,angle=270]{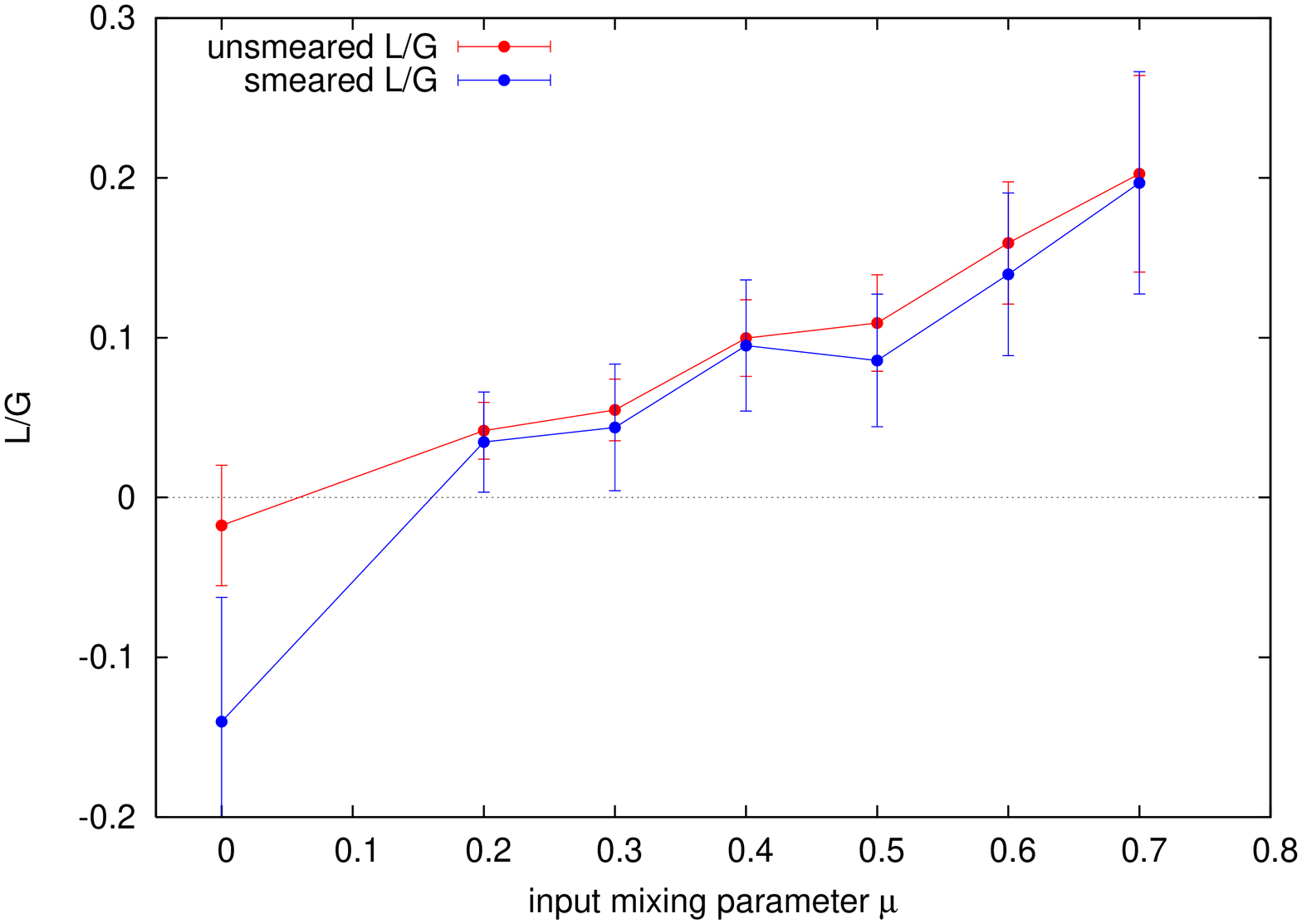}
\caption{Beam shape analysis methods applied to the horizontal projections of simulated Booster beams, as a function of the non-Gaussian component of the beam. The plot on the left shows smeared and unsmeared kurtosis, the plot on the right smeared and unsmeared $L/G$.  The beam width corresponds to a typical Booster horizontal IPM width.} 
\label{fig:resultsx}
\end{figure}

\begin{figure}
\centering
\includegraphics*[height=0.5\columnwidth,angle=270]{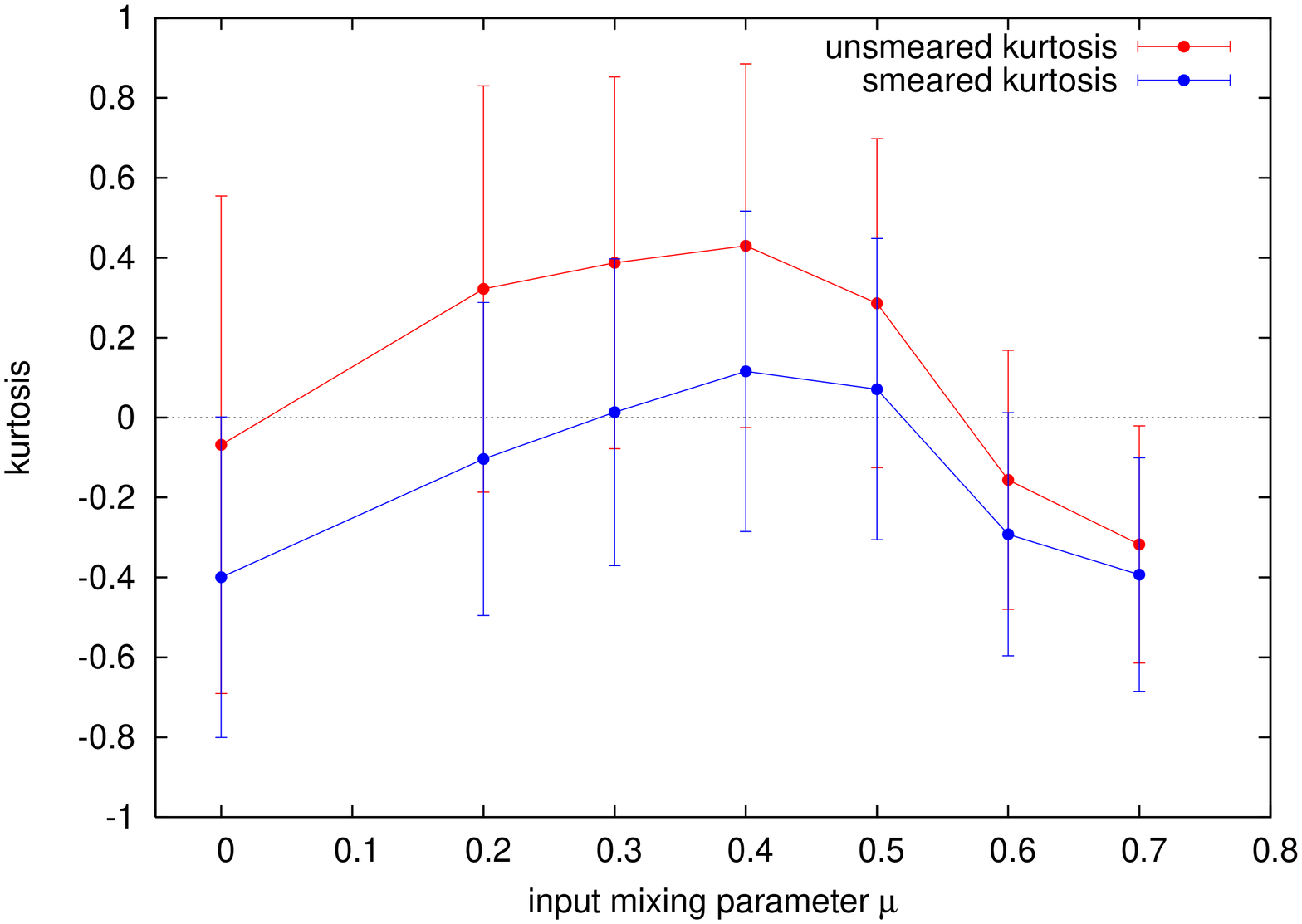}%
\includegraphics*[height=0.5\columnwidth,angle=270]{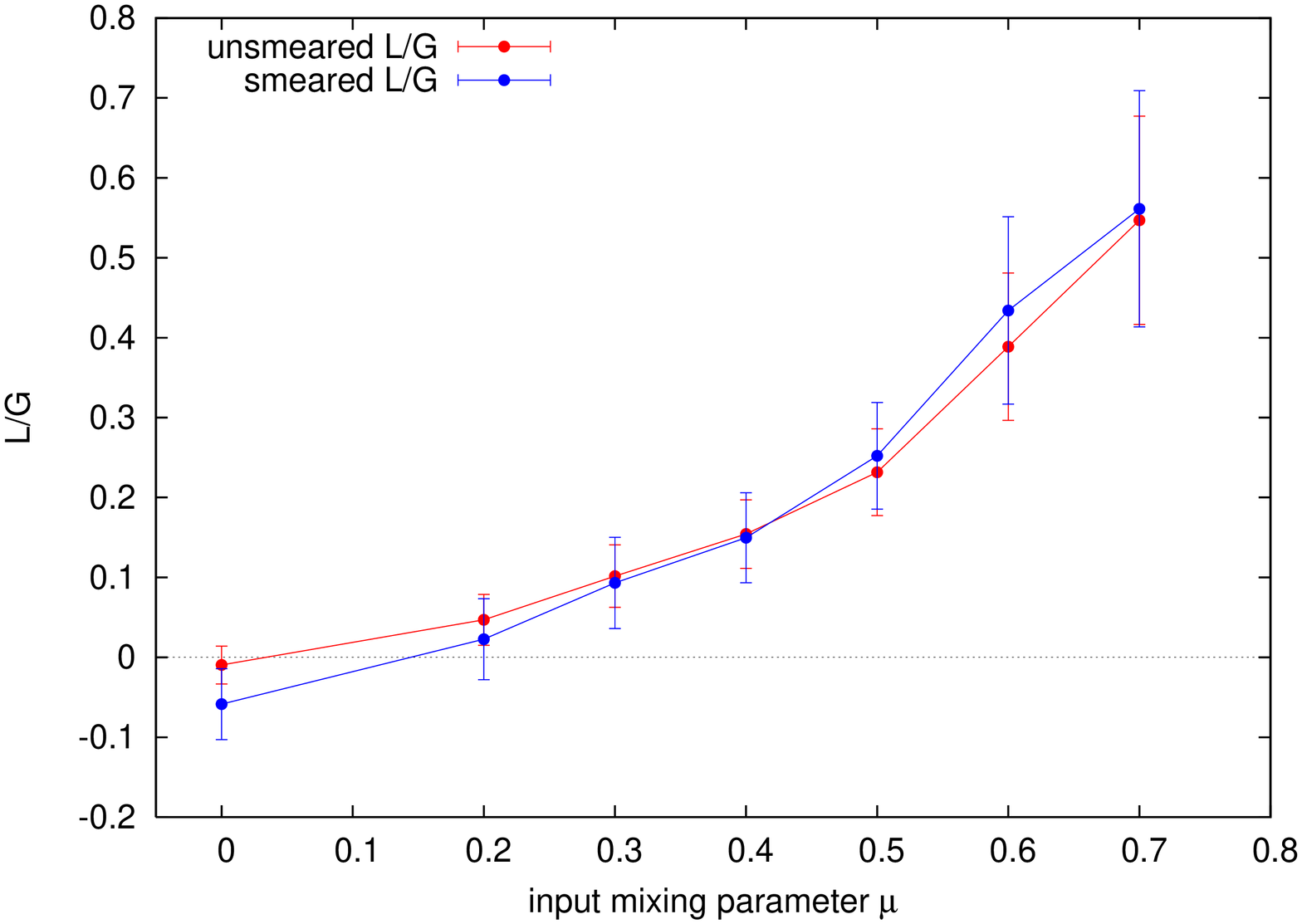}
\caption{Beam shape analysis methods applied to the vertical projections of simulated Booster beams, as a function of the non-Gaussian component of the beam. The plot on the left shows smeared and unsmeared kurtosis, the plot on the right smeared and unsmeared $L/G$.  The beam width corresponds to a typical Booster vertical IPM width.} 
\label{fig:resultsy}
\end{figure}

The results of our simulated studies are summarized in
Figs.~\ref{fig:resultsx} and \ref{fig:resultsy}, where we have calculated kurtosis and $L/G$  from both unsmeared and smeared simulated profiles, as a function of
$\mu$. The two figures correspond to simulated data with different beam widths,
corresponding to the widths shown in the left and right plots of
Figs.~\ref{fig:simu1}--\ref{fig:simu3}. The simulated results in the figures verify our arguments for the superiority of $L/G$ as compared with kurtosis for this analysis: First, $L/G$ increases monotonically with $\mu$, while kurtosis does not. Second, the fractional error bars are much smaller for $L/G$. Third, the effect of detector smearing on $L/G$ is small compared to the overall error, while it is relatively larger for kurtosis. Finally, and most importantly, $L/G$ provides a statistically significant signal for the presence of non-Gaussian beam components, while kurtosis does not.

\section{Beam studies using $L/G$}

As an application of the $L/G$ technique, we have studied the effects
of the beam collimators on the beam shape.  The results of two studies
are presented.  Both studies compare $L/G$ analysis of IPM profile data
for the case where a beam collimator is near the beam, to the case
where the collimator is away from the beam.  One study uses a
collimator located within the Booster ring, while the other uses a
collimator in the linear accelerator (linac).  The
Linac~\cite{ref:linac} is the injector for the Booster synchrotron.

The Fermilab Booster is an alternating gradient synchrotron of radius
75.47 meters. It accelerates protons from 400~MeV to 8 GeV over the
course of 20,000 turns. The optical lattice consists of 24 cells (or
periods) with four combined function magnets each, with horizontal and
vertical tunes of 6.7 and 6.8, respectively. There is a long straight
and short straight section in each period, useful for needed insertion
devices.  The injected beam from the Fermilab Linac has a typical peak
current of 42~mA. The beam is typically injected for ten Booster
turns, for a total average current of 420~mA. The Booster cycles at
15~Hz. A detailed technical description of the Booster can be found in
Ref.~\cite{ref:booster}.

\subsection{Data selection}

For the $L/G$ procedure to work successfully it is very important that
noisy and dead channels are excluded from the fits.  Before analyzing
any set of IPM profiles taken over a particular Booster cycle (an IPM
run), the pedestal information on each IPM channel is used to identify
channels that need to be excluded.  The standard deviation of the
pedestal data for all runs in a given study period is averaged.
Channels with very large standard deviation (noisy), or zero standard
deviation (dead), are excluded from the fits.  Fig.~\ref{fig:peds}
shows the average pedestal sigma (standard deviation) for each IPM
channel for one characteristic IPM data set (35 runs). The figure
shows all the channels recorded by the IPM data acquisition
system. The vertical and horizontal IPM data are encoded in 30
channels each. Channel 31 contains the Booster charge information,
which is why it has a wider pedestal RMS than the rest of the
channels. In this picture, there are no noisy (anomalously large
sigma) channels. However, there are five dead channels, which are
easily identified by their small pedestal variation.

\subsection{$L/G$ with Booster collimators}

The horizontal and vertical collimators within the Booster 
ring are each a two-stage system; a thin copper foil located at a short 
straight section in period 5, followed by secondary collimators in the long 
straight sections of periods 6 and 7.  The beam edge can be put near the 
copper foil in order to scatter halo particles.  The secondary collimators 
pick up the scattered particles~\cite{ref:shielding}.

\begin{figure}[t]
\centering
\includegraphics*[height=0.7\columnwidth,angle=270]{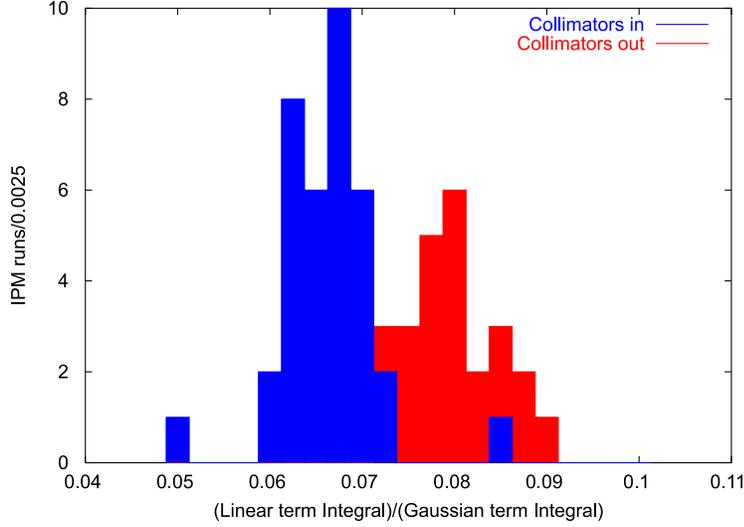}
\caption{Distribution of $L/G$ values in the Booster with and without
the Booster collimators.}
\label{fig:collimators}
\end{figure}

\begin{figure}[t]
\centering
\includegraphics*[width=0.7\columnwidth]{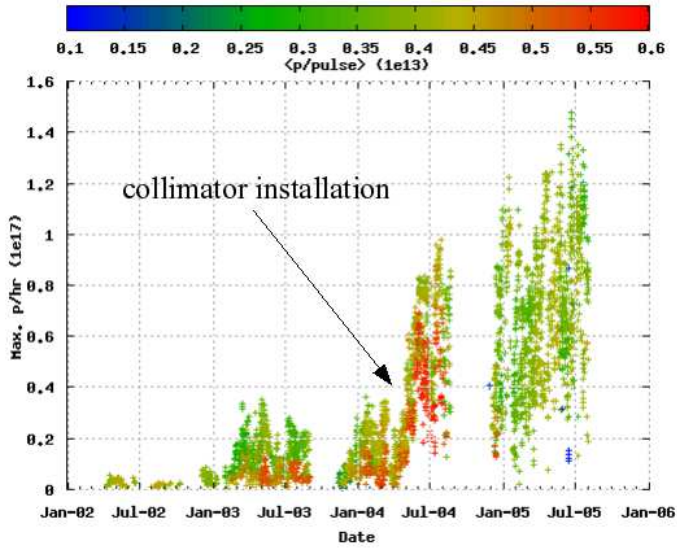}
\caption{Maximum number of protons/hr allowed by the activation limit
in the Booster versus time. The proton per pulse intensity for the
different data sets is shown in different colors. It is clear that the
collimators in the Booster are effective in reducing activation due to
uncontrolled beam loss. The activation limit for Booster throughput
after the installation of the collimators is near 1$\times$10$^{17}$
protons/hr, approximately double the number allowed before the
installation of the collimators.}
\label{fig:pbooster}
\end{figure}

The beam edge is near the collimator only after beam injection is
complete; so, for this study IPM profiles were taken at the end of the
Booster cycle.  Measurements were made both with the collimator in and
with the collimator out, and repeated for several cycles.  The average
value of $L/G$ for 500 turns at the end of the booster cycle was
extracted from each set of profile data (in other words, from each
run).  Fig.~\ref{fig:collimators} shows a histogram of the number of
runs versus their average $L/G$ value.  Even though there is a great
deal of spread in the data, the mean values of $L/G$ for collimators in
and collimators out are distinctly separated.  The overall
distributions clearly show that $L/G$ is lower when the collimators
are in the Booster. Since the Booster collimators are effective in
reducing beam halo (see Fig.~\ref{fig:pbooster}), we conclude that
$L/G$ is a good quantity to use to identify beam halo from beam
profile measurements.

\subsection{$L/G$ with Linac collimator}

The Fermilab Linac accelerates beam from an energy of 750 KeV to 
an energy of 400 MeV.  The Linac collimator is located after the first 
accelerating tank, where the beam energy is 10 MeV.  The collimator has 
several different size apertures that may be inserted into the beam path.  
In order to collimate the beam for these studies, the $1/2$-inch diameter hole 
was used.

\begin{figure}[t]
\centering
\includegraphics*[height=0.7\columnwidth,angle=270]{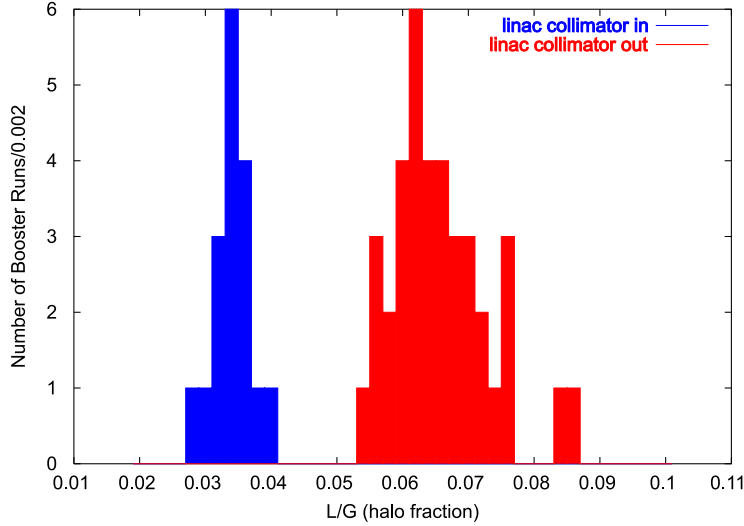}
\caption{Distribution of $L/G$ values in the Booster with and without
the Linac collimators.}
\label{fig:linaccol}
\end{figure}

The procedure was much the same as for the study with the Booster collimator, 
except here the IPM profiles at Booster injection, rather than near 
extraction, were used.  Since the beam halo was reduced in the Linac, the 
effect on the Booster beam profile should have been most pronounced at injection, 
before evolution of the beam distribution through the acceleration cycle.  
IPM profiles for the first 500 turns after beam injection into the Booster 
were used to extract an average $L/G$ value for each run.  Runs were done both with the Linac collimator inserted and with the Linac collimator removed. Fig.~\ref{fig:linaccol} shows a histogram of the number of 
runs versus their average $L/G$ value.  The $L/G$ distributions for collimator-in 
and collimator-out conditions are clearly differentiated.  Presumably the 
distinction is cleaner for the Linac collimator study because the beam 
distribution has not undergone as much evolution between the scraping and 
the profile measurement as in the case of the Booster collimator study.
 
\section{Conclusion}

A new method of characterizing beam halo, the $L/G$ technique, has been developed for beam profile measurements using the Fermilab Booster IPM.  Profile data is fit with a
combination of linear and Gaussian functions.  The ratio of the
integrated values of those functions is taken, with higher $L/G$ values
corresponding to larger tails in the beam distribution.  Simulations
implementing both models of the accelerator and the response of the IPM
show that $L/G$ is superior to kurtosis for characterizing the presence of non-Gaussian portions of the beam both because $L/G$ is a monotonically increasing function of the non-Gaussian fraction of the beam whereas kurtosis is not, and because $L/G$ is less affected by detector errors and smearing than kurtosis.  The $L/G$ quantifier has been tested experimentally in the Booster ring via collimator studies.  Two beam
studies were done, both using the Booster IPM to obtain the needed
profile data.  One study compared injection profile data for the case
of having the Linac beam collimated versus having no collimation.  The
second study compared profile data near extraction for the case of
having the Booster collimator act on the beam versus no collimation.
Both studies demonstrated that the $L/G$ parameter is a good indicator
for beam halo.  We expect that this method will continue to be a
useful tool.


\begin{thebibliography}{9}   

\bibitem{ref:shielding} N.V. Mokhov, A.L. Drozhdin, P.H. Kasper, J.R. Lackey, 
E.J. Prebys, and R.C. Webber, ``Fermilab Booster Beam Collimation and 
Shielding,'' Proceedings of the 2003 Particle Accelerator Conference, 2003.

\bibitem{ref:kurtosis1} T.P. Wangler and K.R. Crandall, ``Beam Halo in Proton 
Linac Beams,'' Linac 2000 Conference Proceedings, Monterey California, 
August 21-25, 2000.

\bibitem{ref:kurtosis2} C.K. Allen and T.P. Wangler, ``Parameters for 
quantifying beam halo,'' Proceedings of the 2001 Particle Accelerator 
Conference, Chicago, 2001.

\bibitem{ref:synergia} J. Amundson, P. Spentzouris, J. Qiang and R. Ryne, 
Journal of Computational Physics, Volume 211, Issue 1, p.~229, 2006.

\bibitem {ref:ipm} J. Zagel, D. Chen, and J. Crisp, {\it Beam
Instrumentation Workshop} (AIP Conference Proceedings 333), p.~384, 1994.

\bibitem{ref:ipmcalibration} J. Amundson, J. Lackey, P. Spentzouris,
G. Jungman, and L. Spentzouris, Phys. Rev. ST Accel. Beams 6:102801, 2003.

\bibitem{allwa} C.K. Allen and T.P. Wangler, Phys. Rev. ST Accel. Beams 5:124202, 2002.

\bibitem{leda} J. Qiang, {\it et al.}, Phys. Rev. ST Accel. Beams 5:124202, 2002. 

\bibitem{ref:booster} Booster Staff, ed.~E.~L.~Hubbard, {\it Booster Synchrotron},
{\it Fermi National Accelerator Laboratory Technical Memo TM-405}, 1973.

\bibitem {ref:linac} C. Ankenbrandt, {\it et al.}, {\it Proceedings of the
11th International Conference on High-Energy Accelerators} p~260, 1980.


\end{thebibliography}
\end{document}